%% file: main.tex
\documentclass[runningheads]{llncs}

\usepackage{graphicx}
\usepackage{amsmath,amssymb}
\usepackage{color}
\usepackage{hyperref}
\hypersetup{
    colorlinks=true,
    linkcolor=blue!70!black,    
    urlcolor=blue!70!black,
    citecolor=blue!70!black,
    filecolor=blue!70!black
}
\DeclareMathOperator*{\argmax}{arg\,max}
\usepackage[table,dvipsnames]{xcolor}
\usepackage{tikz}

\makeatletter
\newcommand\avsuminner[2]{%
  {\sbox0{$\m@th#1\sum$}%
   \vphantom{\usebox0}%
   \ooalign{%
     \hidewidth
     \smash{\vrule height\dimexpr\ht0+1pt\relax depth\dimexpr\dp0+1pt\relax}%
     \hidewidth\cr
     $\m@th#1\sum$\cr
   }%
  }%
}
\makeatother

\definecolor{Dandelion}{RGB}{255, 180, 20}
\definecolor{Purple}{RGB}{160, 32, 240}
\definecolor{RoyalBlue}{RGB}{65, 105, 225}
\definecolor{Gray}{RGB}{190, 190, 190}

\definecolor{gray1}{RGB}{210,210,210}
\definecolor{gray2}{RGB}{50,50,50}
\definecolor{blue2}{RGB}{230,240,250}

\newcommand{\DP}[2]{%
  \begin{tikzpicture}
    \fill[color=#2]   (0.0 , 0.0) rectangle (#1*5.8ex , 2ex );
  \end{tikzpicture}%
}

\usepackage{booktabs}
\usepackage{tabularx}
\usepackage{array}
\usepackage{multirow}       
\usepackage[table,dvipsnames]{xcolor}
\usepackage{colortbl}
\usepackage{arydshln}

\usepackage{esvect}

\usepackage{verbatim}
\usepackage{pifont}

\begin{document}


\title{Data-Efficient Learning for Generalizable Surgical Video Understanding}
\titlerunning{Data-Efficient Learning for Generalizable Surgical Video Understanding}
\author{Sahar Nasirihaghighi}
\authorrunning{Sahar Nasirihaghighi}
\institute{Department of Information Technology (ITEC), University of Klagenfurt\\ \email{Sahar.Nasirihaghighi@aau.at}}

\maketitle
\begin{abstract}
    Advances in surgical video analysis are transforming operating rooms into intelligent, data-driven environments. Computer-assisted systems now support the full surgical workflow, from preoperative planning to intraoperative guidance and postoperative assessment. However, developing robust and generalizable models for surgical video understanding remains challenging due to (I) annotation cost and scarcity, (II) spatiotemporal complexity, and (III) domain gap across procedures and institutions.
    This doctoral research aims to bridge the gap between deep learning–based surgical video analysis in research and its real-world clinical deployment. To address the core challenge of recognizing surgical phases, actions, and events, critical for video-based analysis, I benchmarked state-of-the-art neural network architectures to identify the most effective designs for each task. I further improved performance by proposing novel architectures and integrating advanced modules. Given the high cost of expert annotations and the domain gap across surgical video sources, I focused on reducing reliance on labeled data. We developed semi-supervised frameworks that improve model performance across tasks by leveraging large amounts of unlabeled surgical video.
    We introduced novel semi-supervised frameworks, including DIST, SemiVT-Surge, and ENCORE, that achieved state-of-the-art results on challenging surgical datasets by leveraging minimal labeled data and enhancing model training through dynamic pseudo-labeling.
    To support reproducibility and advance the field, we released two multi-task datasets: GynSurg, the largest gynecologic laparoscopy dataset, and Cataract-1K, the largest cataract surgery video dataset.
    Together, this work contributes to robust, data-efficient, and clinically scalable solutions for surgical video analysis, laying the foundation for generalizable AI systems that can meaningfully impact surgical care and training.
    
\end{abstract}

\input{1-Research_Problem_and_Motivation}
\input{2-Background}
\input{3-Scientific_Approach}

\input{4-Proposed_Solution}

\input{5-Results_and_Contribution}

\input{6-Open_Challenges_and_Future_Work}
\input{7-Long-Term_Goals}

\bibliographystyle{splncs04}
\bibliography{mybibliography}

\end{document}

%% file: 1-Research_Problem_and_Motivation.tex
\section{Research Problem and Motivation}
Operating rooms are rapidly evolving into intelligent, data-driven environments, driven by the integration of advanced technologies. Central to this transformation are computer-assisted systems (CAS), which support the entire surgical workflow, from preoperative planning and intraoperative decision-making to automated skill assessment and postoperative analysis~\cite{guan2024label,ma2022simulation}. A cornerstone of these systems is the ability to automatically recognize surgical workflows and understand the operating scene, enabling critical capabilities such as procedural indexing, summarization, performance evaluation, and report generation~\cite{ghamsarian2020enabling}. This transformation is further accelerated by the increasing availability of surgical videos, which provide rich data for developing intelligent models aimed at improving the safety, efficiency, and quality of surgical care~\cite{guan2024label}. In this context, deep learning has emerged as a key enabler, powering the automated recognition of surgical phases and actions, as well as the semantic segmentation of anatomical structures and surgical instruments.

Despite the growing availability of surgical video data, effectively leveraging this resource remains a major research challenge~\cite{liu2025lovit}. Surgical videos are inherently complex, with significant variability across procedures, patients, and clinical environments. A central obstacle is the scarcity and imbalance of annotated data, as producing high-quality labels requires significant time and clinical expertise, making large-scale annotation both costly and limited in scope~\cite{dave2023timebalance,liu2024most}. This is further compounded by visual artifacts such as rapid instrument motion, occlusions, poor lighting, smoke, and blood, which degrade video quality and hinder automated analysis.
Another major barrier to model generalization is the domain gap between datasets. Variations in imaging devices, compression methods, patient demographics, and surgical techniques across hospitals or even individual operating rooms introduce substantial visual discrepancies. These inconsistencies often prevent models trained in one domain from performing reliably in another, limiting their real-world applicability. Given these challenges, this research is guided by the following key questions:

\begin{itemize}
\item How can we reduce reliance on large-scale expert annotations without compromising model performance?
\item How can we leverage abundant available unlabeled surgical video data to improve intra-domain performance and mitigate cross-domain domain shifts?
\item How can we design data-efficient, temporally-aware, and generalizable models that perform reliably across diverse surgical settings?
\end{itemize}

\begin{figure}[!t]
    \centering
    \includegraphics[width=1\textwidth]{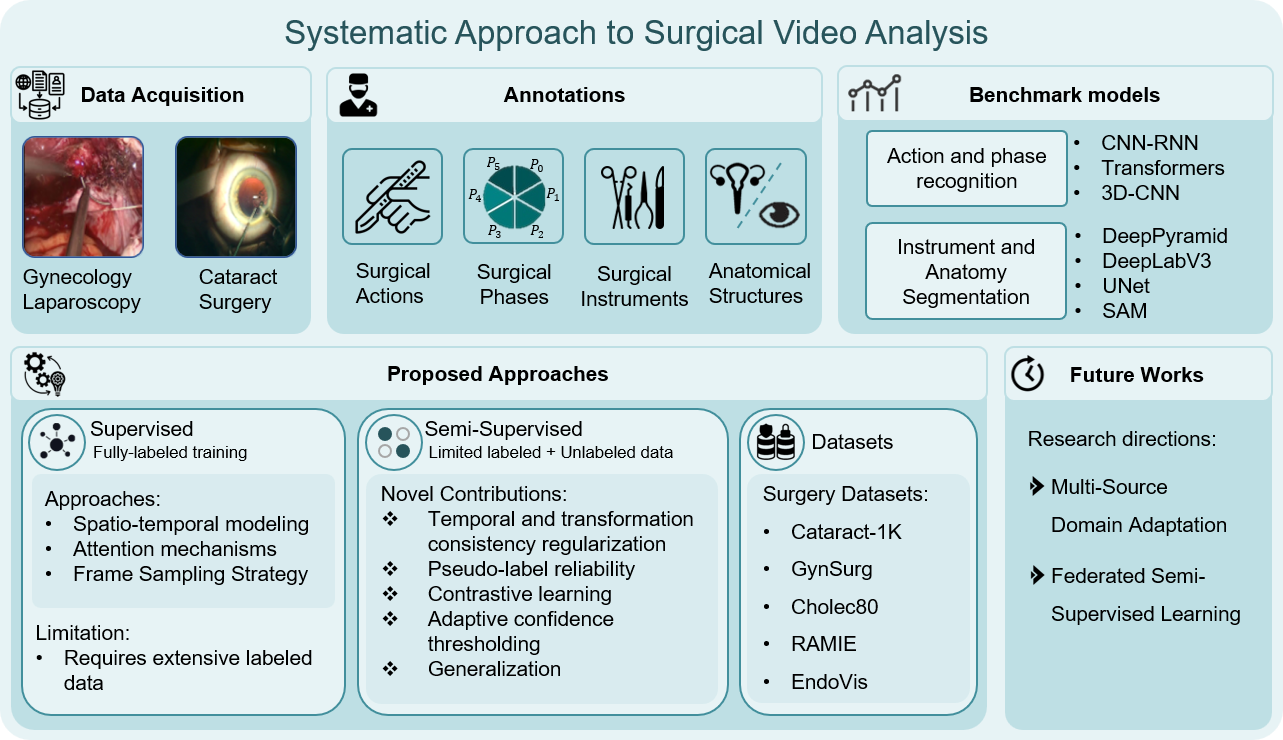}
    \caption{Multi-Stage Pipeline for Scalable Surgical Video Analysis}
    \label{fig:pipeline}
\end{figure}

To answer these questions, this work focuses on developing robust and scalable deep learning frameworks for surgical video understanding. The proposed approaches integrate supervised and semi-supervised learning strategies to reduce annotation demands, employ temporal modeling to capture procedural dynamics, and incorporate domain-aware training to improve generalization across datasets. 
Recognizing that data acquisition, curation, and annotation are essential for enabling real-world surgical video analysis, we paid careful attention to these aspects in our research. As part of our contributions to the community, we curated and publicly released two large-scale, multi-task datasets: GynSurg~\cite{nasirihaghighi2025gynsurg} and Cataract-1K~\cite{ghamsarian2024cataract}. In addition to these datasets, this research encompasses multiple surgical video and imaging domains, focusing on key tasks such as action/event recognition, surgical phase recognition, and semantic segmentation of instruments and relevant anatomical structures. Figure~\ref{fig:pipeline} illustrates the overall research pipeline, outlining the full workflow from data acquisition and annotation to model benchmarking, proposed learning approaches, and future research directions.
\vspace{-0.5em}

%% file: 2-Background.tex
\section{Background}
This section provides a comprehensive overview of the latest advancements in surgical video analysis. Given the present study’s focus on phase and action recognition and instrument and anatomy segmentation, we aim to highlight methods in this domain on surgical videos.

\paragraph{Action and Phase Recognition.}
Understanding surgical workflows through video analysis is a foundational task in computer-assisted interventions, enabling both intraoperative guidance and postoperative assessment. Initial approaches typically used image-level deep models that analyzed frames independently, but these failed to capture the temporal continuity of surgical procedures. To address this limitation, hybrid architectures combining spatial encoders with temporal sequence models such as CNN-RNNs were introduced, significantly improving performance~\cite{ghamsarian2021relevance}. For example, TeCNO by Czempiel et al.~\cite{czempiel2020tecno} uses a multi-stage Temporal Convolutional Network (TCN) to enhance temporal consistency and phase recognition. Although supervised methods perform well in controlled settings, their generalization across surgical domains is limited due to reliance on large annotated datasets and sensitivity to domain shifts across institutions or procedures. To address these limitations, semi-supervised learning (SSL) has gained traction by leveraging large amounts of unlabeled surgical data alongside limited annotations. Among the explored strategies, three main paradigms have shown particular promise: contrastive learning~\cite{singh2021semi,lou2023min}, consistency regularization~\cite{ke2020guided,varsavsky2020test}, and pseudo-supervision~\cite{feng2022dmt,wei2021crest}. These approaches help reduce annotation costs while maintaining strong performance in surgical phase recognition.

\paragraph{Semantic Segmentation.}
Pixel-level identification of anatomical structures and surgical instruments plays a critical role in surgical video analysis. To this end, UNet-based encoder-decoder architectures have demonstrated strong performance in fully supervised settings~\cite{ghamsarian2024deeppyramid+}. To alleviate the need for extensive annotations, semi-supervised learning techniques such as entropy minimization, consistency regularization, and teacher-student frameworks have also been explored for semantic segmentation. For instance, Su et al.\cite{su2024mutual} assess pseudo-label reliability based on intra-class feature similarity, whereas  BCP~\cite{bai2023bidirectional} mitigates domain mismatch through cross-domain augmentation strategies.

\paragraph{Shortcomings of Existing Methods.}
Despite significant progress in surgical video analysis, existing methods face several critical limitations. In tasks such as phase and action recognition, current supervised methods tend to overlook temporal variation and contextual continuity, which are essential for accurate and robust performance. While semi-supervised methods reduce labeling needs, they often rely on simplistic pseudo-labeling and fixed thresholds, leading to noisy supervision in low-label settings. These challenges underscore the need for data-efficient, temporally-aware, and robust semi-supervised-learning frameworks, an area that forms the core focus of this research.

%% file: 3-Scientific_Approach.tex
\section{Scientific Approach}

\begin{figure}[!t]
    \centering
    \includegraphics[width=1\textwidth]{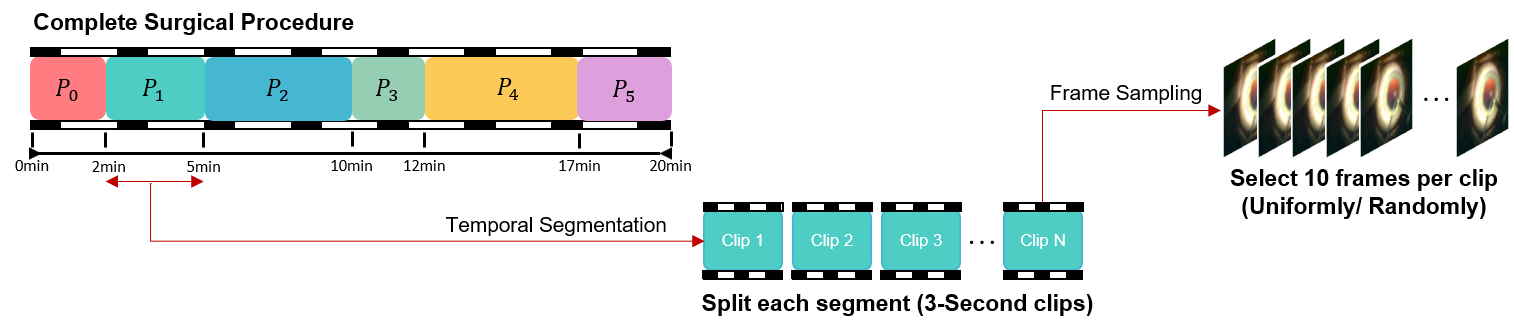}
    \caption{\small Frame sampling strategy}
    \label{fig:frame_sampling}
\end{figure}

To address key challenges in surgical video understanding, including annotation scarcity, temporal ambiguity, and cross-domain variability, we developed a series of supervised and semi-supervised learning frameworks for action recognition, phase classification, and semantic segmentation across multiple domains, including gynecologic laparoscopy and cataract surgery.

\paragraph{\textbf{Dataset Curation.}}To support semantic segmentation in surgical video analysis, we curated and publicly released two of the largest and most comprehensive datasets to date: \textit{GynSurg}\cite{nasirihaghighi2025gynsurg}, focused on gynecologic laparoscopy, and \textit{Cataract-1K}\cite{ghamsarian2024cataract}, focused on cataract surgery. Each dataset includes multi-task annotations encompassing surgical phases, actions, and pixel-level segmentations of surgical instruments and anatomical structures. These datasets were designed to address the unique visual and procedural characteristics of their respective domains and to facilitate benchmarking in both intra- and cross-domain settings.

\paragraph{\textbf{Supervised Action and Phase Recognition.}} 
Our early work focused on fine-grained action recognition in gynecologic laparoscopic surgery, a domain marked by complex visual conditions such as smoke, bleeding, occlusion, and rapid camera motion. We proposed a CNN-RNN hybrid architecture that integrates spatial feature extractors with stacked recurrent layers to capture temporal dependencies critical for recognizing short and variable actions~\cite{nasirihaghighi2023action}. To improve robustness to surgeon-specific variability such as differences in speed, technique, and temporal execution, we introduced a frame sampling strategy that enhances temporal diversity during training as illustrated in Fig.~\ref{fig:frame_sampling}.
Building on this foundation, we addressed the detection of critical intraoperative events such as bleeding. Although gynecologic laparoscopy is minimally invasive, intraoperative bleeding remains a safety-critical event that can affect surgical outcomes. To better model long-range temporal dependencies, we extended our earlier architecture by integrating transformer encoders into a hybrid CNN-transformer framework~\cite{nasirihaghighi2024event}. Spatial features extracted by were passed through multi-head self-attention layers, enabling the model to recognize temporally complex events including abdominal access, bleeding, needle passing, and coagulation or transection.

\paragraph{\textbf{Supervised Semantic Segmentation.}}We benchmarked a diverse set of segmentation models to evaluate intra-domain and cross-domain performance across the curated datasets. The models span classic encoder-decoder architectures such as UNet and DeepLabV3, pyramid-based designs like DeepPyramid, and transformer-based foundation models such as Segment Anything (SAM), which we fine-tuned using LoRA and prompt-based adaptation strategies. Intra-domain performance was assessed by training and testing within the same dataset, while cross-domain generalization was evaluated by training on Cataract-1K and testing on CaDIS~\cite{grammatikopoulou2021cadis} using binary instrument segmentation. This comprehensive benchmarking provides a standardized reference for future supervised segmentation models in surgical video analysis.

\paragraph{\textbf{Semi-Supervised Learning.}} 
While our supervised models significantly improved performance, they rely heavily on large volumes of expert-annotated data. To reduce this dependency, we adopted semi-supervised learning (SSL), which combines limited labeled data with abundant unlabeled data to improve generalization and mitigate overfitting. We explored SSL methods across different tasks, developing tailored strategies for both classification and segmentation settings. These include temporal consistency regularization, confidence-based pseudo-label filtering, and contrastive learning to improve representation learning in low-label regimes.

\begin{figure}[!t]
    \centering
    \includegraphics[width=1\textwidth]{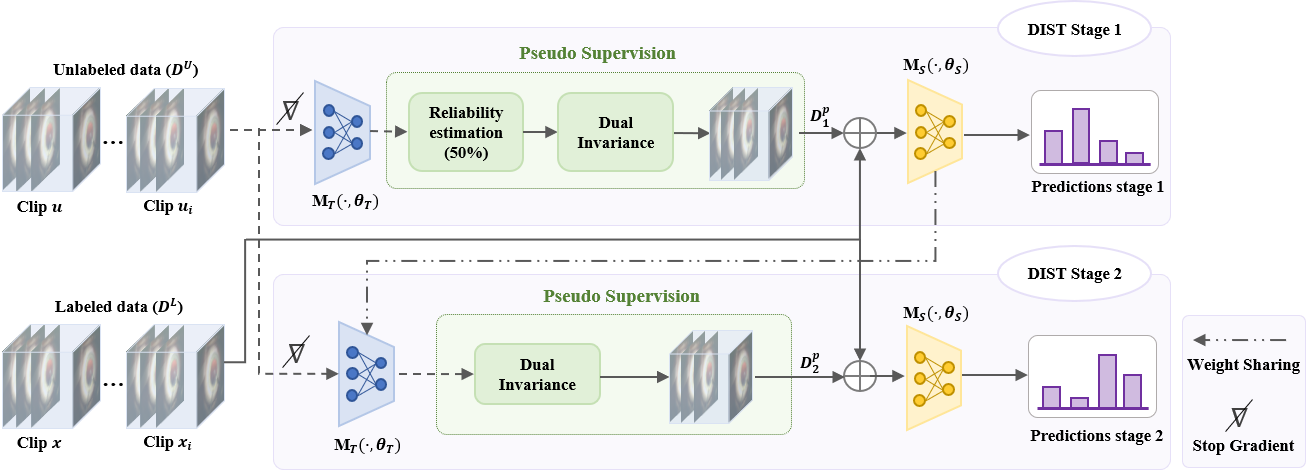}
    \caption{\small Dual Invariance Self-training framework}
    \label{fig:DIST_framework}
\end{figure}

%% file: 4-Proposed_Solution.tex
\section{Proposed Solution}

Building on the earlier introduction, this section outlines the core methodologies developed for semi-supervised surgical video analysis, emphasizing the learning strategies and innovations across action and phase recognition and semantic segmentation.

\paragraph{\textbf{Dual Invariance Self-Training for Surgical Phase Recognition.}} We introduced \textit{Dual Invariance Self-Training (DIST)}~\cite{nasirihaghighi2025dual}, a two-stage semi-supervised framework for action and phase recognition. DIST aims to improve pseudo-label quality using a dual filtering mechanism. In Stage 1, a teacher model trained on labeled data generates pseudo-labels for unlabeled clips. Reliability is estimated via temporal consistency, using predictions from three training checkpoints: \(T(n/3)\), \(T(2n/3)\), and \(T(n)\). For each unlabeled clip \(u_i\), we compute:

\vspace{-1em}
\begin{align}
z_i^n = & \sigma(\mathcal{M}_{T}(u_i^f, \theta_T(n))) \notag \\ 
\hat{z}_i^n = &\argmax_C(z_i^n)
\end{align}
\vspace{-1.2em}
\begin{equation}
\mathcal{R}_{u_i} = \Big(\frac{z_i^{\frac{n}{3}} \times \hat{z}_i^{n}}{z_i^{\frac{n}{3}} + \hat{z}_i^{n}} + 2\times\frac{z_i^{\frac{2n}{3}} \times \hat{z}_i^{n}}{z_i^{\frac{2n}{3}} + \hat{z}_i^{n}}\Big)/3
\end{equation}

\noindent Only the top $50\%$ of consistent pseudo-labels are retained.
To further filter unreliable predictions, DIST enforces two invariance constraints: temporal invariance (across different frame sampling strategies) and transformation invariance (under strong augmentations). Only labels satisfying both conditions are used to train a student model. In Stage 2, the student becomes the teacher, and the process is repeated.
Evaluated on Cataract-1K and Cholec80, DIST outperforms both supervised and state-of-the-art SSL methods, especially in low-label scenarios. Its dual filtering strategy significantly improves pseudo-label robustness and generalization. In Stage 1, a teacher model generates pseudo-labels, which are filtered by a reliability estimation technique based on prediction consistency across three checkpoints.

\paragraph{\textbf{Semi-Supervised Video Transformer for Surgical Phase Recognition.}}
We further proposed \textit{SemiVT-Surg}~\cite{li2025semivt}, a transformer-based SSL framework for surgical phase recognition. It incorporates two components: (1) \textbf{Contrastive Learning with Class Prototypes (CLP)}, which aligns embeddings with class centroids while pushing them away from negative prototypes, and (2) \textbf{Temporal Consistency Regularization (TCR)}, which enforces consistency between augmented clip predictions in a dual-stream teacher-student setup. Trained on RAMIE and Cholec80 using the TimeSformer backbone, the model achieved strong cross-domain generalization and superior temporal reasoning over CNN-based baselines.
The CLP module uses a triplet loss to structure the feature space, pulling embeddings toward their class prototype while repelling them from hard negative prototypes, both of which are updated online via exponential moving average. TCR leverages long-short term frame sampling and applies confidence thresholding to filter noisy pseudo-labels, improving training stability and reducing confirmation bias. Extensive ablation studies show that both components contribute complementary gains, and integrating a causal TCN module further improves phase boundary modeling.

\paragraph{\textbf{Ensemble-of-Confidence Reinforcement for Semantic Segmentation.}}  
We proposed ENCORE~\cite{ghamsarian2025feedback}, a feedback-driven pseudo-label selection method that replaces fixed thresholding with an adaptive confidence estimation pipeline. ENCORE introduces two main components:  
(1) \textbf{Class-Aware Confidence Calibration (CAC)}, which estimates class-wise thresholds using the distribution of true-positive confidence scores in labeled data; and  
(2) \textbf{Adaptive Confidence Thresholding (ACT)}, which employs assessor networks to evaluate threshold candidates and selects the optimal one based on real-time Dice performance.
By dynamically adjusting thresholds based on model feedback, ENCORE eliminates the need for manual tuning or large validation sets. It improves both training stability and label quality, and consistently outperforms baseline SSL segmentation methods in low-label scenarios across EndoVis, Cataract-1K, and other datasets.

%% file: 5-Results_and_Contribution.tex
\input{Table/DIST_results}
\section{Results and Contribution}

Our research involved training and evaluating models on a diverse set of datasets, including GynLap6, GynSurg, Cataract-1K, RAMIE, Cholec80, and EndoVis, covering tasks such as action and phase recognition, as well as instrument and anatomy segmentation. This broad dataset coverage allowed us to validate performance across a wide range of surgical scenarios and imaging modalities, demonstrating the robustness and adaptability of our approaches.

Our initial work with CNN-RNN~\cite{nasirihaghighi2023action} and transformer-based architectures~\cite{nasirihaghighi2024event} achieved strong performance in recognizing surgical actions and critical events, such as bleeding, in gynecologic laparoscopic videos, significantly outperforming frame-based methods. 
The proposed DIST framework consistently outperforms state-of-the-art SSL methods such as FixMatch~\cite{sohn2020fixmatch}, Mean Teacher~\cite{liu2022perturbed}, Cross Pseudo Supervision~\cite{CMPL}, and SVFormer~\cite{svformer}. As shown in Table~\ref{tab:DIST}, DIST achieves strong results on challenging low-label splits (e.g., 85.97\% accuracy with only one labeled video in the 1/32 split) on both Cataract-1K and Cholec80. We further validated DIST across different architectures and conducted ablation studies to assess the impact of dual invariance filtering (see~\cite{nasirihaghighi2025dual} for details). Figure~\ref{tab:phase-prediction} illustrates qualitative comparisons on a Cataract-1K video case.
The SemiVT-Surge model~\cite{li2025semivt}, built on the TimeSformer architecture, was trained on the private RAMIE dataset and Cholec80. It achieved an accuracy of 83.4\% on RAMIE, outperforming both supervised and existing semi-supervised baselines. Ablation studies confirmed that both Temporal Consistency Regularization (TCR) and Contrastive Learning with Prototypes (CLP) contributed significantly to the model's performance gains.

Figure~\ref{fig:Lap_benchmark_models} shows the F1-scores of action recognition models benchmarked in the GynSurg study~\cite{nasirihaghighi2025gynsurg}. In addition to action recognition, the dataset includes annotations for side effects such as bleeding and smoke, enabling a more comprehensive analysis of intraoperative conditions. We also benchmarked multiple segmentation models for identifying surgical instruments, auxiliary tools, and anatomical structures in gynecologic laparoscopy. The Segment Anything Model (SAM) was fine-tuned using both grid and point prompts, with the point-based approach achieving the best performance across all categories, demonstrating SAM’s strong adaptability and precise localization in surgical environments.

In the ENCORE study \cite{ghamsarian2025feedback}, the framework was evaluated across different datasets, including Cataract-1K and EndoVis, using segmentation models such as UNet, VNet, and DeepLabV3+. ENCORE consistently improved Dice scores and model robustness, particularly in low-label regimes. It is fully compatible with existing semi-supervised frameworks like UniMatch, Switch, and AD-MT, providing immediate gains without requiring architectural modifications. Table~\ref{tab:ENCORE} summarizes ENCORE’s performance improvements over state-of-the-art baselines on the most challenging low-label splits.

\input{Table/DIST_phase_prediction}

%% file: Table/DIST_results.tex
\begin{table}[t!]
\centering
\label{tab:side_effect_recognition}
\begin{minipage}{0.49\linewidth}
\centering
\caption{\small Evaluation of the DIST framework based on Accuracy and F1-score (\%).}

\label{tab:DIST}
\resizebox{\linewidth}{!}{%

\begin{tabular}{>{\raggedright\arraybackslash}m{2.2cm} m{2cm} *{4}{>{\centering\arraybackslash}m{1.5cm}}}

\specialrule{.12em}{.05em}{.05em}
 & & \multicolumn{2}{c}{\textbf{1/16}} & \multicolumn{2}{c}{\textbf{1/32}} \\ \cmidrule(lr){3-4}\cmidrule(lr){5-6}
Dataset & Model & Acc. & F1 & Acc. & F1 \\ \specialrule{.12em}{.05em}{.05em}
\multirow{4}{*}{Cataract-1K}
& CMPL & 83.26 & 79.87 & 72.47 & 60.98 \\
& SVFormer & 84.04 & 80.38 & 73.55 & 67.14 \\
& DIST\_Stage1 & 91.22& 90.16  & 85.93  & 84.02\\
& DIST\_Stage2 & 92.78 & 91.60& 85.97 & 83.62 \\
\end{tabular}
}
\resizebox{\linewidth}{!}{%
\begin{tabular}{>{\raggedright\arraybackslash}m{2.2cm} m{2cm} *{4}{>{\centering\arraybackslash}m{1.5cm}}}

\specialrule{.12em}{.05em}{.05em}
 & & \multicolumn{2}{c}{\textbf{1/16}} & \multicolumn{2}{c}{\textbf{1/32}} \\ \cmidrule(lr){3-4}\cmidrule(lr){5-6}
Dataset & Model & Acc. & F1 & Acc. & F1 \\ \specialrule{.12em}{.05em}{.05em}
\multirow{4}{*}{Cholec80}
& SVFormer & 75.84 & 70.21 & 64.26 & 60.51 \\
& DIST\_Stage1 & 79.52 & 79.43 & 74.63  & 73.44\\
& DIST\_Stage2 & 79.95 & 79.97& 75.46 & 74.31 \\
\specialrule{.12em}{.05em}{.05em}
\end{tabular}
}

\end{minipage}
\hfill
\begin{minipage}{0.49\linewidth}
\centering
\caption{\small Evaluation of the ENCORE framework based on dice coefficient (\%).}
\label{tab:ENCORE}
\resizebox{\linewidth}{!}{%
\begin{tabular}{>{\raggedright\arraybackslash}m{2cm} m{3.5cm} *{2}{>{\centering\arraybackslash}m{1.8cm}}}
\specialrule{.12em}{.05em}{.05em}
Dataset & Model & 1/8 & 1/26 \\ \specialrule{.12em}{.05em}{.05em}
\multirow{2}{*}{Cataract-1K}
& AD-MT & 78.46 & 62.74 \\
& AD-MT+ENCORE & 84.33 & 71.62 \\
\end{tabular}
}
\resizebox{\linewidth}{!}{%
\begin{tabular}{>{\raggedright\arraybackslash}m{2cm} m{3.5cm} *{2}{>{\centering\arraybackslash}m{1.8cm}}}
\specialrule{.12em}{.05em}{.05em}
Dataset & Model & 1/8 & 1/26 \\ \specialrule{.12em}{.05em}{.05em}
\multirow{2}{*}{Cataract-1K}
& Switch & 81.24 & 62.81 \\
& Switch+ENCORE & 89.39 & 83.21 \\

\end{tabular}
}
\resizebox{\linewidth}{!}{%
\begin{tabular}{>{\raggedright\arraybackslash}m{2cm} m{3.5cm} *{2}{>{\centering\arraybackslash}m{1.8cm}}}
\specialrule{.12em}{.05em}{.05em}
Dataset & Model & 1/8 & 1/16 \\ \specialrule{.12em}{.05em}{.05em}
\multirow{2}{*}{EndoVis}
& UniMatch & 78.42  & 77.34\\
& UniMatch+ENCORE & 80.93  & 79.54 \\

\specialrule{.12em}{.05em}{.05em}
\end{tabular}
}
\end{minipage}
\end{table}

%% file: Table/DIST_phase_prediction.tex
\begin{figure}[!t]
\centering
\begin{minipage}{0.52\textwidth}
    \centering
    \caption{\small Phase prediction performance of various models using the hybrid transformer network on the Cataract-1k dataset.}
    \vspace{0.5em}
    \label{tab:phase-prediction}
    \resizebox{\linewidth}{!}{%
    \begin{tabular}{lll}
    \specialrule{.12em}{.05em}{.05em}
    Split & Model & Phases\\\midrule
    \multicolumn{2}{c}{Ground-Truth} & \DP{2.00}{Gray}\DP{0.18}{Gray}\DP{3.04}{RoyalBlue}\DP{0.25}{Gray}\DP{1.17}{Dandelion}\DP{0.09}{Gray}\DP{0.48}{Gray}\DP{0.09}{Gray}\DP{0.27}{Purple}\DP{0.11}{Gray}\DP{0.23}{Gray}\DP{0.06}{Gray}\DP{0.50}{Dandelion}\DP{0.01}{Gray}\DP{0.11}{Dandelion}\DP{0.07}{Gray}\DP{0.40}{Gray}\DP{0.05}{Gray}\DP{0.25}{Dandelion}\DP{0.05}{Gray}\DP{0.59}{Gray}\\\hdashline
    \multirow{4}{*}{1/4}&Supervised& \DP{2.04}{Gray}\DP{0.00}{Gray}\DP{0.08}{RoyalBlue}\DP{0.00}{Gray}\DP{0.08}{Purple}\DP{0.00}{Gray}\DP{0.23}{Gray}\DP{0.00}{Gray}\DP{0.08}{Dandelion}\DP{0.00}{Gray}\DP{0.08}{Gray}\DP{0.00}{Gray}\DP{0.15}{Dandelion}\DP{0.00}{Gray}\DP{2.49}{RoyalBlue}\DP{0.00}{Gray}\DP{0.30}{Gray}\DP{0.00}{Gray}\DP{1.13}{Dandelion}\DP{0.00}{Gray}\DP{0.75}{Gray}\DP{0.00}{Gray}\DP{0.15}{Purple}\DP{0.00}{Gray}\DP{0.45}{Gray}\DP{0.00}{Gray}\DP{0.60}{Dandelion}\DP{0.00}{Gray}\DP{0.53}{Gray}\DP{0.00}{Gray}\DP{0.30}{Dandelion}\DP{0.00}{Gray}\DP{0.53}{Gray}\DP{0.00}{Gray}\DP{0.04}{Purple}\\
    &MeanTeacher&\DP{1.96}{Gray}\DP{0.00}{Gray}\DP{0.15}{Dandelion}\DP{0.00}{Gray}\DP{3.09}{RoyalBlue}\DP{0.00}{Gray}\DP{0.30}{Gray}\DP{0.00}{Gray}\DP{1.13}{Dandelion}\DP{0.00}{Gray}\DP{0.75}{Gray}\DP{0.00}{Gray}\DP{0.23}{Purple}\DP{0.00}{Gray}\DP{0.38}{Gray}\DP{0.00}{Gray}\DP{0.60}{Dandelion}\DP{0.00}{Gray}\DP{0.53}{Gray}\DP{0.00}{Gray}\DP{0.30}{Dandelion}\DP{0.00}{Gray}\DP{0.57}{Gray}\\
    &SVFormer&\DP{2.15}{Gray}\DP{0.00}{Gray}\DP{0.15}{Dandelion}\DP{0.00}{Gray}\DP{0.12}{Gray}\DP{0.00}{Gray}\DP{0.08}{Dandelion}\DP{0.00}{Gray}\DP{0.15}{Gray}\DP{0.00}{Gray}\DP{0.13}{RoyalBlue}\DP{0.00}{Gray}\DP{0.03}{Dandelion}\DP{0.00}{Gray}\DP{0.38}{RoyalBlue}\DP{0.00}{Gray}\DP{0.03}{Dandelion}\DP{0.00}{Gray}\DP{1.75}{RoyalBlue}\DP{0.00}{Gray}\DP{0.03}{Dandelion}\DP{0.00}{Gray}\DP{0.15}{RoyalBlue}\DP{0.00}{Gray}\DP{0.05}{Dandelion}\DP{0.00}{Gray}\DP{0.28}{Gray}\DP{0.00}{Gray}\DP{0.93}{Dandelion}\DP{0.00}{Gray}\DP{0.03}{RoyalBlue}\DP{0.00}{Gray}\DP{0.20}{Dandelion}\DP{0.00}{Gray}\DP{0.73}{Gray}\DP{0.00}{Gray}\DP{0.15}{Purple}\DP{0.00}{Gray}\DP{0.40}{Gray}\DP{0.00}{Gray}\DP{0.55}{Dandelion}\DP{0.00}{Gray}\DP{0.03}{RoyalBlue}\DP{0.00}{Gray}\DP{0.07}{Dandelion}\DP{0.00}{Gray}\DP{0.55}{Gray}\DP{0.00}{Gray}\DP{0.23}{Dandelion}\DP{0.00}{Gray}\DP{0.68}{Gray}\\
    &DIST&\DP{2.24}{Gray}\DP{0.00}{Gray}\DP{2.97}{RoyalBlue}\DP{0.00}{Gray}\DP{0.28}{Gray}\DP{0.00}{Gray}\DP{1.16}{Dandelion}\DP{0.00}{Gray}\DP{0.70}{Gray}\DP{0.00}{Gray}\DP{0.23}{Purple}\DP{0.00}{Gray}\DP{0.40}{Gray}\DP{0.00}{Gray}\DP{0.60}{Dandelion}\DP{0.00}{Gray}\DP{0.55}{Gray}\DP{0.00}{Gray}\DP{0.23}{Dandelion}\DP{0.00}{Gray}\DP{0.64}{Gray}\\
    \hdashline
    \multirow{4}{*}{1/16}&Supervised&\DP{2.04}{Gray}\DP{0.00}{Gray}\DP{0.23}{Dandelion}\DP{0.00}{Gray}\DP{0.15}{RoyalBlue}\DP{0.00}{Gray}\DP{0.53}{Dandelion}\DP{1.15}{Gray}\DP{0.21}{RoyalBlue}\DP{0.00}{Gray}\DP{0.08}{Dandelion}\DP{0.00}{Gray}\DP{0.38}{RoyalBlue}\DP{0.00}{Gray}\DP{0.08}{Dandelion}\DP{0.00}{Gray}\DP{0.38}{RoyalBlue}\DP{0.00}{Gray}\DP{0.08}{Dandelion}\DP{0.00}{Gray}\DP{0.23}{Gray}\DP{0.00}{Gray}\DP{0.83}{Dandelion}\DP{0.00}{Gray}\DP{0.15}{RoyalBlue}\DP{0.00}{Gray}\DP{0.15}{Dandelion}\DP{0.00}{Gray}\DP{0.75}{Gray}\DP{0.00}{Gray}\DP{0.15}{Purple}\DP{0.00}{Gray}\DP{0.45}{Gray}\DP{0.00}{Gray}\DP{0.60}{Dandelion}\DP{0.00}{Gray}\DP{0.53}{Gray}\DP{0.00}{Gray}\DP{0.30}{Dandelion}\DP{0.00}{Gray}\DP{0.53}{Gray}\DP{0.00}{Gray}\DP{0.04}{Purple}\\
    &MeanTeacher&\DP{2.72}{Gray}\DP{0.00}{Gray}\DP{2.49}{RoyalBlue}\DP{0.00}{Gray}\DP{0.38}{Gray}\DP{0.00}{Gray}\DP{0.53}{Dandelion}\DP{0.00}{Gray}\DP{0.08}{Gray}\DP{0.00}{Gray}\DP{0.15}{Dandelion}\DP{0.00}{Gray}\DP{0.15}{Gray}\DP{0.00}{Gray}\DP{0.15}{Dandelion}\DP{0.00}{Gray}\DP{1.43}{Gray}\DP{0.00}{Gray}\DP{0.53}{Dandelion}\DP{0.00}{Gray}\DP{0.60}{Gray}\DP{0.00}{Gray}\DP{0.23}{Dandelion}\DP{0.00}{Gray}\DP{0.57}{Gray}\\
    &SVFormer&\DP{1.95}{Gray}\DP{0.00}{Gray}\DP{0.03}{Dandelion}\DP{0.00}{Gray}\DP{0.05}{Gray}\DP{0.00}{Gray}\DP{0.05}{Dandelion}\DP{0.00}{Gray}\DP{0.05}{Gray}\DP{0.00}{Gray}\DP{0.05}{Dandelion}\DP{0.00}{Gray}\DP{0.50}{Gray}\DP{0.00}{Gray}\DP{0.08}{Dandelion}\DP{0.00}{Gray}\DP{0.13}{Gray}\DP{0.00}{Gray}\DP{0.18}{Dandelion}\DP{0.00}{Gray}\DP{0.08}{RoyalBlue}\DP{0.00}{Gray}\DP{0.10}{Gray}\DP{0.00}{Gray}\DP{1.95}{RoyalBlue}\DP{0.00}{Gray}\DP{0.30}{Gray}\DP{0.00}{Gray}\DP{1.13}{Dandelion}\DP{0.00}{Gray}\DP{0.73}{Gray}\DP{0.00}{Gray}\DP{0.18}{Dandelion}\DP{0.00}{Gray}\DP{0.38}{Gray}\DP{0.00}{Gray}\DP{0.65}{Dandelion}\DP{0.00}{Gray}\DP{0.55}{Gray}\DP{0.00}{Gray}\DP{0.23}{Dandelion}\DP{0.00}{Gray}\DP{0.68}{Gray}\\
    &DIST&\DP{2.59}{Gray}\DP{0.00}{Gray}\DP{0.08}{Dandelion}\DP{0.00}{Gray}\DP{2.54}{RoyalBlue}\DP{0.00}{Gray}\DP{0.28}{Gray}\DP{0.00}{Gray}\DP{1.16}{Dandelion}\DP{0.00}{Gray}\DP{0.73}{Gray}\DP{0.00}{Gray}\DP{0.23}{Purple}\DP{0.00}{Gray}\DP{0.38}{Gray}\DP{0.00}{Gray}\DP{0.60}{Dandelion}\DP{0.00}{Gray}\DP{0.55}{Gray}\DP{0.00}{Gray}\DP{0.23}{Dandelion}\DP{0.00}{Gray}\DP{0.64}{Gray}\\
    \specialrule{.12em}{.05em}{.05em}
    \end{tabular}
    }
    \resizebox{\linewidth}{!}{%
    \begin{tabular}{m{1.5cm}m{8cm}}
    Colormap: &  Irrigation-Aspiration \DP{0.9}{Dandelion}, 
    Lens Implantation \DP{0.9}{Purple},
    Phako \DP{0.9}{RoyalBlue},
    Rest \DP{0.9}{Gray}
    \\
    \specialrule{.12em}{.05em}{.05em}
    \end{tabular}
    }
   
\end{minipage}%
\hfill
\begin{minipage}{0.45\textwidth}
    \centering
    \includegraphics[width=\linewidth]{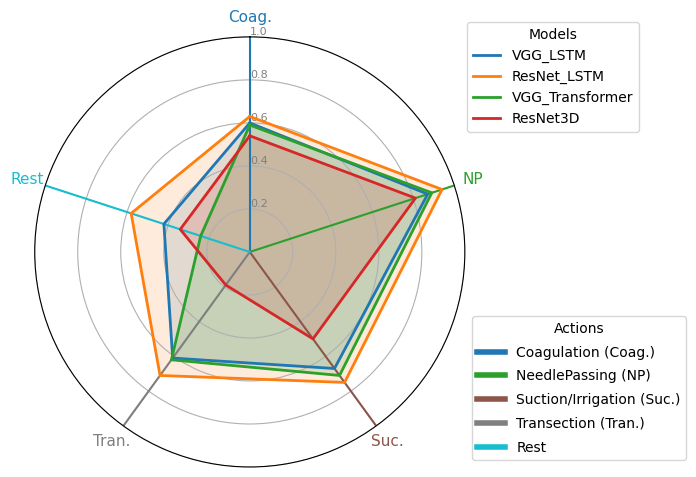}
    \caption{\small F1-scores for each action across different models on GynSurg dataset.}
    \label{fig:Lap_benchmark_models}
\end{minipage}
\end{figure}

%% file: 6-Open_Challenges_and_Future_Work.tex
\section{Open Challenges and Future Work}
\vspace{-0.5em}
While my research has advanced supervised and semi-supervised surgical video analysis, clinical deployment challenges remain. Domain shift--caused by variability in imaging systems, protocols, and environments across institutions--limits model generalization in unseen settings. I am investigating domain adaptation and federated learning to enhance robustness without centralized data sharing.
Federated semi-supervised learning is particularly valuable in healthcare, where labeled data is scarce and privacy regulations prohibit patient data transfer. This approach enables decentralized training across hospitals by leveraging local unlabeled data alongside small shared labeled datasets. It preserves patient confidentiality while improving generalization across heterogeneous sources and supports continuous learning on evolving clinical data, providing a practical solution for scalable, privacy-aware deployment in surgical AI systems.

\vspace{-0.5em}

%% file: 7-Long-Term_Goals.tex
\section{Long Term Goals}
\vspace{-0.5em}
My research focuses on advancing surgical video understanding through supervised and semi-supervised learning across gynecologic laparoscopy and cataract surgery. My vision is to develop clinically deployable, generalizable, real-time AI systems that strengthen surgical decision-making, improve patient safety, and enhance surgical training. I intend to pursue an academic research career bridging deep learning and clinical practice through interdisciplinary collaboration, translating technological innovations into tangible clinical impact.
\vspace{-0.5em}

%% file: main.bbl
\begin{thebibliography}{10}
\providecommand{\url}[1]{\texttt{#1}}
\providecommand{\urlprefix}{URL }
\providecommand{\doi}[1]{https://doi.org/#1}

\bibitem{bai2023bidirectional}
Bai, Y., Chen, D., Li, Q., Shen, W., Wang, Y.: Bidirectional copy-paste for semi-supervised medical image segmentation. In: Proceedings of the IEEE/CVF conference on computer vision and pattern recognition. pp. 11514--11524 (2023)

\bibitem{czempiel2020tecno}
Czempiel, T., Paschali, M., Keicher, M., Simson, W., Feussner, H., Kim, S.T., Navab, N.: Tecno: Surgical phase recognition with multi-stage temporal convolutional networks. In: International conference on medical image computing and computer-assisted intervention. pp. 343--352. Springer (2020)

\bibitem{dave2023timebalance}
Dave, I.R., Rizve, M.N., Chen, C., Shah, M.: Timebalance: Temporally-invariant and temporally-distinctive video representations for semi-supervised action recognition. In: Proceedings of the IEEE/CVF Conference on Computer Vision and Pattern Recognition. pp. 2341--2352 (2023)

\bibitem{feng2022dmt}
Feng, Z., Zhou, Q., Gu, Q., Tan, X., Cheng, G., Lu, X., Shi, J., Ma, L.: Dmt: Dynamic mutual training for semi-supervised learning. Pattern Recognition  (2022)

\bibitem{ghamsarian2020enabling}
Ghamsarian, N.: Enabling relevance-based exploration of cataract videos. In: Proceedings of the 2020 int. conf. on multimedia retrieval. pp. 378--382 (2020)

\bibitem{ghamsarian2024cataract}
Ghamsarian, N., El-Shabrawi, Y., Nasirihaghighi, S., Putzgruber-Adamitsch, D., Zinkernagel, M., Wolf, S., Schoeffmann, K., Sznitman, R.: Cataract-1k dataset for deep-learning-assisted analysis of cataract surgery videos. Scientific data  \textbf{11}(1), ~373 (2024)

\bibitem{ghamsarian2025feedback}
Ghamsarian, N., Nasirihaghighi, S., Schoeffmann, K., Sznitman, R.: Feedback-driven pseudo-label reliability assessment: Redefining thresholding for semi-supervised semantic segmentation. arXiv preprint arXiv:2505.07691  (2025)

\bibitem{ghamsarian2021relevance}
Ghamsarian, N., Taschwer, M., Putzgruber-Adamitsch, D., Sarny, S., Schoeffmann, K.: Relevance detection in cataract surgery videos by spatio-temporal action localization. In: 2020 25th International conference on pattern recognition (ICPR). pp. 10720--10727. IEEE (2021)

\bibitem{ghamsarian2024deeppyramid+}
Ghamsarian, N., Wolf, S., Zinkernagel, M., Schoeffmann, K., Sznitman, R.: Deeppyramid+: medical image segmentation using pyramid view fusion and deformable pyramid reception. International journal of computer assisted radiology and surgery  \textbf{19}(5),  851--859 (2024)

\bibitem{grammatikopoulou2021cadis}
Grammatikopoulou, M., Flouty, E., Kadkhodamohammadi, A., Quellec, G., Chow, A., Nehme, J., Luengo, I., Stoyanov, D.: Cadis: Cataract dataset for surgical rgb-image segmentation. Medical Image Analysis  \textbf{71},  102053 (2021)

\bibitem{guan2024label}
Guan, J., Zou, X., Tao, R., Zheng, G.: Label-guided teacher for surgical phase recognition via knowledge distillation. In: International Conference on Medical Image Computing and Computer-Assisted Intervention. pp. 349--358. Springer (2024)

\bibitem{ke2020guided}
Ke, Z., Qiu, D., Li, K., Yan, Q., Lau, R.W.: Guided collaborative training for pixel-wise semi-supervised learning. In: European conference on computer vision. pp. 429--445. Springer (2020)

\bibitem{li2025semivt}
Li, Y., de~Jong, R., Nasirihaghighi, S., Jaspers, T., van Jaarsveld, R., Kuiper, G., van Hillegersberg, R., van~der Sommen, F., Ruurda, J., Breeuwer, M., et~al.: Semivt-surge: Semi-supervised video transformer for surgical phase recognition. arXiv preprint arXiv:2506.01471  (2025)

\bibitem{liu2024most}
Liu, X., Chen, Z., Yuan, Y.: Most: multi-formation soft masking for semi-supervised medical image segmentation. In: International Conference on Medical Image Computing and Computer-Assisted Intervention. pp. 469--480. Springer (2024)

\bibitem{liu2025lovit}
Liu, Y., Boels, M., Garcia-Peraza-Herrera, L.C., Vercauteren, T., Dasgupta, P., Granados, A., Ourselin, S.: Lovit: Long video transformer for surgical phase recognition. Medical Image Analysis  \textbf{99},  103366 (2025)

\bibitem{liu2022perturbed}
Liu, Y., Tian, Y., Chen, Y., Liu, F., Belagiannis, V., Carneiro, G.: Perturbed and strict mean teachers for semi-supervised semantic segmentation. In: Proceedings of the IEEE/CVF conf. on computer vision and pattern recognition (2022)

\bibitem{lou2023min}
Lou, A., Tawfik, K., Yao, X., Liu, Z., Noble, J.: Min-max similarity: A contrastive semi-supervised deep learning network for surgical tools segmentation. IEEE Transactions on Medical Imaging  \textbf{42}(10),  2832--2841 (2023)

\bibitem{ma2022simulation}
Ma, L., Xiao, D., Kim, D., Lian, C., Kuang, T., Liu, Q., Deng, H., Yang, E., Liebschner, M.A., Gateno, J., et~al.: Simulation of postoperative facial appearances via geometric deep learning for efficient orthognathic surgical planning. IEEE transactions on medical imaging  \textbf{42}(2),  336--345 (2022)

\bibitem{nasirihaghighi2024event}
Nasirihaghighi, S., Ghamsarian, N., Husslein, H., Schoeffmann, K.: Event recognition in laparoscopic gynecology videos with hybrid transformers. In: International Conference on Multimedia Modeling. pp. 82--95. Springer (2024)

\bibitem{nasirihaghighi2025gynsurg}
Nasirihaghighi, S., Ghamsarian, N., Peschek, L., Munari, M., Husslein, H., Sznitman, R., Schoeffmann, K.: Gynsurg: A comprehensive gynecology laparoscopic surgery dataset. arXiv preprint arXiv:2506.11356  (2025)

\bibitem{nasirihaghighi2023action}
Nasirihaghighi, S., Ghamsarian, N., Stefanics, D., Schoeffmann, K., Husslein, H.: Action recognition in video recordings from gynecologic laparoscopy. In: 2023 IEEE 36th International symposium on computer-based medical systems (CBMS). pp. 29--34. IEEE (2023)

\bibitem{nasirihaghighi2025dual}
Nasirihaghighi, S., Ghamsarian, N., Sznitman, R., Schoeffmann, K.: Dual invariance self-training for reliable semi-supervised surgical phase recognition. In: 2025 IEEE 22nd Int. Symp. on Biomedical Imaging (ISBI). pp. 01--05. IEEE (2025)

\bibitem{singh2021semi}
Singh, A., Chakraborty, O., Varshney, A., Panda, R., Feris, R., Saenko, K., Das, A.: Semi-supervised action recognition with temporal contrastive learning. In: Proceedings of the IEEE/CVF conference on computer vision and pattern recognition. pp. 10389--10399 (2021)

\bibitem{sohn2020fixmatch}
Sohn, K., Berthelot, D., Carlini, N., Zhang, Z., Zhang, H., Raffel, C.A., Cubuk, E.D., Kurakin, A., Li, C.L.: Fixmatch: Simplifying semi-supervised learning with consistency and confidence. Advances in neural information processing system  \textbf{33},  596--608 (2020)

\bibitem{su2024mutual}
Su, J., Luo, Z., Lian, S., Lin, D., Li, S.: Mutual learning with reliable pseudo label for semi-supervised medical image segmentation. Medical Image Analysis  \textbf{94},  103111 (2024)

\bibitem{varsavsky2020test}
Varsavsky, T., Orbes-Arteaga, M., Sudre, C.H., Graham, M.S., Nachev, P., Cardoso, M.J.: Test-time unsupervised domain adaptation. In: Int. Conf. on Medical Image Computing and Computer-Assisted Intervention. Springer (2020)

\bibitem{wei2021crest}
Wei, C., Sohn, K., Mellina, C., Yuille, A., Yang, F.: Crest: A class-rebalancing self-training framework for imbalanced semi-supervised learning. In: Proceedings of the IEEE/CVF conference on computer vision and pattern recognition. pp. 10857--10866 (2021)

\bibitem{svformer}
Xing, Z., Dai, Q., Hu, H., Chen, J., Wu, Z., Jiang, Y.G.: Svformer: Semi-supervised video transformer for action recognition. In: Proceedings of the IEEE/CVF conference on computer vision and pattern recognition. pp. 18816--18826 (2023)

\bibitem{CMPL}
Xu, Y., Wei, F., Sun, X., Yang, C., Shen, Y., Dai, B., Zhou, B., Lin, S.: Cross-model pseudo-labeling for semi-supervised action recognition. In: Proceedings of the IEEE/CVF Conference on Computer Vision and Pattern Recognition. pp. 2959--2968 (2022)

\end{thebibliography}
